\documentstyle[12pt,epsf]{article}
\renewcommand{\thefootnote}{\fnsymbol{footnote}}
\begin{document}
\footnotetext[1]{The complete paper, including figures, is
also available via anonymous ftp at
ttpux2.physik.uni-karlsruhe.de (129.13.102.139) as /ttp94-24/ttp94-24.ps,
or via www at http://ttpux2.physik.uni-karlsruhe.de/preprints.html/}
\title{Polarization in Top Pair Production\\
and Decay near Threshold
%
\footnote[2]{Presented by M. Je\.zabek at
XI International Symposium on High
Energy Spin Physics, SPIN'94,
Sept.15-22, 1994, Bloomington, Indiana, USA.}
}
\author{
R. Harlander$^a$,
M. Je\.zabek$^{a,b}$,
J.H. K\"uhn$^a$ and T. Teubner$^a$\\
{\normalsize \it $^a$ Institut f\"ur Theoretische Teilchenphysik,
D-76128 Karlsruhe, Germany}\\
{\normalsize \it $^{b}$ Institute of Nuclear Physics,
Kawiory 26a, PL-30055 Cracow, Poland}
}
\date{}
\maketitle
\thispagestyle{empty}
\vspace{-3.0in}
\begin{flushright}
{\bf TTP94-24\footnotemark[1]}\\
{\bf November 1994}
\end{flushright}
\vspace{2.5in}
\begin{abstract}
\noindent
{\small
Theoretical results are presented for top quarks
produced in annihilation of polarized electrons
on positrons.  Polarization studies for $t\bar t$
pairs near threshold are free from hadronization ambiguities.
This is due to the short lifetime of the top quark.
Semileptonic decays are discussed as well as their
applications in studying polarization  dependent processes
involving top quarks.
The Green function formalism is applied to $t\bar t$
production at future $e^+e^-$ colliders with polarized beams.
Lippmann--Schwinger equation is solved numerically for the QCD
chromostatic potential given by the two-loop formula
at large momentum transfers and Richardson ansatz
at intermediate and small ones. The polarization dependent
momentum distributions of top quarks and their decay products
are calculated.
}
\end{abstract}
%
\renewcommand{\thefootnote}{\arabic{footnote}}
\section{Introduction}
The top quark is the heaviest fermion of the Standard Model.
Its large mass allows to probe deeply into the QCD potential
for nonrelativistic $t\bar t$ system produced near energy
threshold. Such a system will provide a unique opportunity
for a variety of novel QCD studies. The lifetime
of the top quark is shorter than the formation time
of top mesons and toponium resonances. Therefore top decays
intercept the process of hadronization at an early stage
and practically eliminate associated nonperturbative effects.

The analysis of polarized top quarks and their decays
has recently attracted considerable attention,
see \cite{Kuehn3,teupitz} and references
cited therein.
The reason is that this analysis
will result in determination of the top
quark coupling to the $W$ and $Z$ bosons either
confirming the predictions
of the Standard Model
or providing clues for physics beyond.
The latter possibility is particularly intriguing
because $m_t$
plays an exceptional role in the fermion mass spectrum.

The polarization fourvector $s^\mu$
of the top quark can be determined
from the angular-energy distributions of the charged leptons
in semileptonic $t$ decays.
In the $t$ quark rest frame this distribution
is in Born approximation
the product of the energy
and the angular distributions\cite{JK89b}:
\begin{equation}
{ {\rm d}^2 \Gamma\over{\rm d}E_\ell\,{\rm d}\cos\theta} =
{1\over 2}\, \left[\, 1\,+
\,S\cos\theta \right]\,{{\rm d}\Gamma\over{\rm d}E_\ell}
\label{eq:elec1}
\end{equation}
where $s^\mu=(0,\vec s\,)$, $S= |\,\vec s\,|$
and $\theta$ is the
angle between $\vec s$ and the direction of the charged lepton.
QCD corrections essentially do not spoil factorization\cite{CJK91}.
Thus, the polarization analyzing power of
the charged lepton energy-angular distribution
remains maximal.
There is no factorization
for the neutrino energy-angular distribution
which is therefore less sensitive to the
polarization of the decaying top quark.
On the other hand it has been shown \cite{JK94} that
the angular-energy distribution of neutrinos
from the polarized top quark decay will allow for a particularly
sensitive test of the V-A structure of the weak charged current.

A number of mechanisms has been suggested that will
lead to polarized top quarks. However, studies at
a linear electron-positron collider are particularly
clean for precision tests.
Moreover, close to threshold
and with longitudinally polarized electrons
one can study decays of polarized top quarks
under particularly convenient conditions:
large event rates, well identified rest frame of the top quark,
and large degree of polarization.
At the same time,
thanks to the spectacular success
of the polarization program at SLC \cite{woods},
the longitudinal polarization of the electron beam
will be an obvious option
for a future linear collider\footnote{Another proposed
and closely related
facility is a photon linear collider.
At such a machine the high energy photon beams
can be generated via Compton scattering of laser
light on  electrons accelerated in the linac.
The threshold behaviour of the reaction
$\gamma\gamma\to t\bar t$
has been reviewed in \cite{DESYWac} and the
top quark polarization
has been recently considered in \cite{fkk}.}.
In this article some results are
presented of a recent calculation \cite{hjkt} of
top quark polarization
for the reaction $e^+e^-\to t\bar t$ in the threshold
region.
In Sect.2 we discuss the dependence of
the top quark polarization  on the longitudinal
polarizations of the beams.
Due to restricted phase space the amplitude is
dominantly $S$ wave and the electron and positron
polarizations are directly transferred to the top quark.
For a quantitative study
this simple picture
has to be extended and the modifications
originating from $S-P$ wave interference
should be taken into account.
In Sect.3 all these corrections are calculated
from numerical solutions of Lippmann-Schwinger
equations.

\section{Top quark polarization}
We adopt the conventions of ref.\cite{alexan}
and describe
the longitudinal polarization of the $e^+e^-$ system
in its center-of-mass frame as a function of the variable
\begin{equation}
\chi = {P_{e^+}-P_{e^-}\over 1 - P_{e^+}P_{e^-}}
\end{equation}
where $P_{{e^\pm}}$ denote the polarizations of $e^\pm$
with respect to the directions of $e^+$ and $e^-$ beams,
respectively\footnote{It is conceivable
that for a future linear $e^+e^-$ collider
$P_{e^+}=0$, $P_{e^-}\ne 0$ and then $\chi = -P_{e^-}$.}.
In the absence of phases from final state interaction,
which can be induced by higher orders in $\alpha_s$ and will be
considered elsewhere \cite{hjkt}, the top quark polarization
is in the production plane.
$P_{^\|}$ and $P_{\bot}$ denote
the longitudinal and the transverse
components of top polarization vector (in its rest frame)
with respect to the electron beam. The angle
$\vartheta$ denotes the angle between $e^-$ and
the top quark.
In the threshold region the top quark is nonrelativistic (with
velocity $\beta=p/m_t\sim\alpha_s$)
and the kinetic energy of the $t\bar t$ system
$E=\sqrt{s}-2m_t$ is of the order ${\cal O}(\beta^2)$.
Retaining only the terms up to
${\cal O}(\beta)$ one derives the following expressions for
the components of the polarization vector:
\begin{eqnarray}
P_{^\|} &=& C^0_{^\|}(\chi) + C^1_{^\|}(\chi)\Phi(E)\cos\vartheta
\label{eq:2}\\
P_\bot &=&C_\bot(\chi)\Phi(E)\sin\vartheta
\label{eq:3}
\end{eqnarray}
\vskip -0.5cm
\begin{figure}[htb]
\epsfxsize=5.5truein
\leavevmode
\epsffile[30 300 525 515]{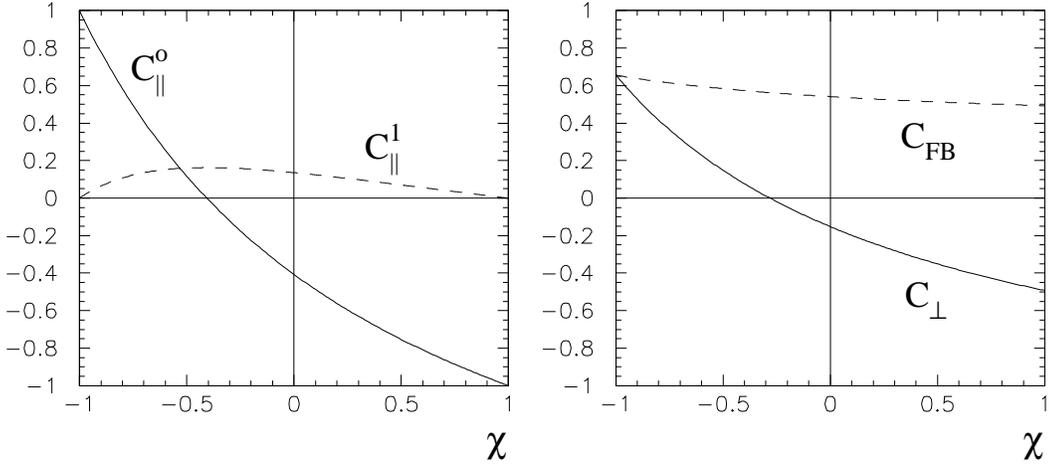}
\vskip-0.2cm
\caption{Coefficient functions: a) $C^0_{^\|}(\chi)$ -- solid line
and $C^1_{^\|}(\chi)$ -- dashed line, b) $C_\bot(\chi)$ -- solid line
and $C_{FB}(\chi)$ -- dashed line.}
\label{fig-CLchi}
\end{figure}

\noindent
The coefficients $C^0_{^\|}(\chi)$, $C^1_{^\|}(\chi)$
and $C_\bot(\chi)$ depend on the polarization $\chi$, the
electroweak coupling constants, the $Z$ mass and the center-of-mass
energy $\sqrt{s}\approx 2m_t$. They are plotted in
Fig.\ref{fig-CLchi} for $m_t=174$~GeV.
$C^0_{^\|}(\chi)$ and $C^1_{^\|}(\chi)$
are shown in Fig.\ref{fig-CLchi}a as the solid and the dashed
lines, respectively, and $C_\bot(\chi)$ as the solid line in
Fig.\ref{fig-CLchi}b.

\noindent
The function
$\Phi(E)$ describes the complicated dynamics of the
$t\bar t$ system near threshold. In particular
it includes effects of the would-be toponium resonances
and Coulomb enhancement. Nevertheless, it is possible to calculate
this function using the Green function method.
The same function
$\Phi(E)$ also governs the
forward-backward asymmetry
in $e^+e^-\to t\bar t$
\begin{equation}
A_{FB} = C_{FB}(\chi)\Phi(E)
\label{eq:4}
\end{equation}
where $C_{FB}$
is shown as the dashed line in Fig.\ref{fig-CLchi}b.
Eqs.(\ref{eq:2}) and (\ref{eq:3})
extend the results of \cite{krz} into the threshold region.
\section{Lippmann-Schwinger equations}
\begin{figure}[htb]
\begin{center}
\epsfxsize=4.5truein
\vskip-0.5cm
\leavevmode
\epsffile[70 125 500 700]{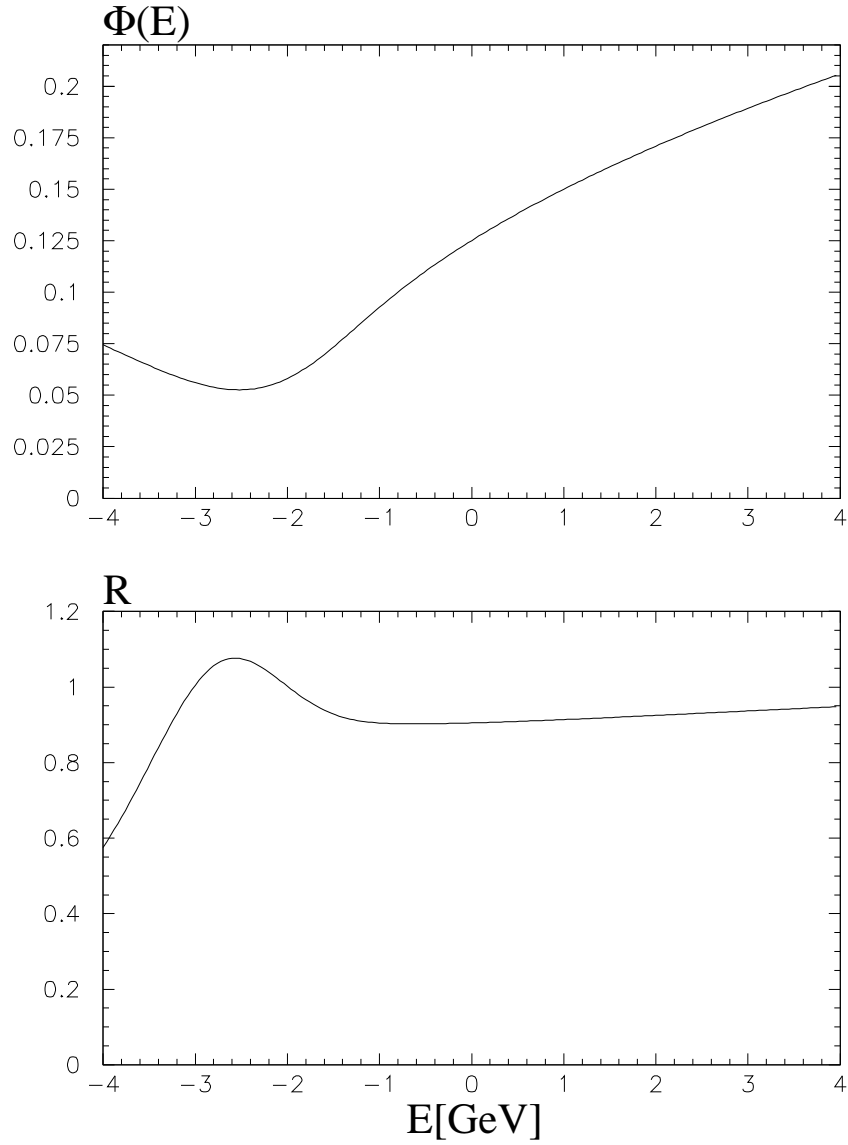}
\end{center}
\vskip-0.8cm
\caption{Energy dependence in the threshold region
of: a) $\Phi(E)$ and
b) $R=\sigma(e^+e^-\to t\bar t\,)/\sigma(e^+e^-\to \mu^+\mu^-)$
for $m_t=174$~GeV and $\alpha_s(m_Z)=0.12$.}
\label{fig-FiR}
\end{figure}

The Green function method has become a standard tool for studying
$e^+e^-$ annihilation in the threshold region
\cite{FK,SP,Sumino1,JKT}. We follow the momentum space approach
of \cite{JKT} and solve the Lippmann-Schwinger
equations numerically for the $S$-wave and $P$-wave Green functions
\begin{eqnarray}
G(p,E) &=&
G_0(p,E) +
G_0(p,E)
\int {{\rm d}^3q\over(2\pi)^3}
\tilde V\left(\,|\,\vec{p}-\vec{q}\,|\,\right)
G(q,E)
\label{eq:LS1}\\
F(p,E) &=&
G_0(p,E) +
G_0(p,E)
\int {{\rm d}^3q\over(2\pi)^3}
{{\vec p}\cdot{\vec q}\over p^2}
\tilde V\left(\,|\,\vec{p}-\vec{q}\,|\,\right)
F(q,E)
\label{eq:LS2}
\end{eqnarray}
where $p=|\,\vec p\,|$ is the momentum
of the top quark in $t\bar t$ rest frame,
\begin{equation}
G_0(p,E) = \left(\,
E- {p^2/ m_t}+ {\rm i}\Gamma_t\,\right)^{-1}\,.
\label{eq:8}
\end{equation}
$\Gamma_t$ denotes the top width
and $\tilde V(p)$ is the QCD potential in momentum
space; see \cite{JKT,hjkt} for details. The function $\Phi(E)$
is related to $G(p,E)$ and $F(p,E)$:
\begin{equation}
\Phi(E) = {
\left( 1- {4\alpha_s\over 3\pi} \right) {1\over m_t}
\int_0^{p_{m}}\,{\rm d}p\, p^3 {\cal R}e \left( \,G\,F^*\,\right)
\over
\left( 1- {8\alpha_s\over 3\pi} \right)
\int_0^{p_{m}}\,{\rm d}p\, p^2 \left|\, G\,\right|^2  }
\label{eq:Phi}
\end{equation}
where $p_{m}$ has been introduced in order to cut off
a logarithmic divergence of the numerator.
The denominator remains finite for $p_{m}\rightarrow\infty$.
In experimental analyses the
contributions of very large intrinsic momenta
will be automatically suppressed
by separation of $t\bar t$ events from the background.
In our calculation we use $p_{m}= m_t/3$. The function
$\Phi(E)$ is plotted in Fig.\ref{fig-FiR}a for $m_t=174$~GeV.
The QCD potential depends on $\alpha_s(m_Z)$ and our results
have been obtained for $\alpha_s=0.12$. For a comparison in
Fig.\ref{fig-FiR}b the annihilation cross section
$\sigma(e^+e^-\to t\bar t\,)$
is shown in units of the annihilation cross section
$\sigma(e^+e^-\to \mu^+\mu^-)$.
\pagebreak[4]

\section*{Acknowledgments}
MJ would like to thank
the Stefan Batory Foundation for a travel grant
and the Local Organizing Committee
for an additional support which
enabled his participation in SPIN'94.
This work was supported in part
by KBN grant 2P30225206 and by DFG contract 436POL173193S.
\end{document}